\documentclass[prb,amsmath,amssymb,reprint,showpacs]{revtex4-1}
\usepackage{graphicx}
\usepackage{dcolumn}
\usepackage{bm}
\newcommand{\p}{$\%$}
\newcommand{\pat}{$\mathrm{~at.}\%~ $}
\newcommand{\che}[1]{$\mathrm{#1}$}

\newcommand{\gt}{$\gamma^{\prime\prime\prime}$}

\newcommand{\pn}{$R\mathrm{_{N_2}}$}
\newcommand{\FeN}{Fe-N}
\newcommand{\FeTi}{$\mathrm{Fe_{0.98}Ti_{0.02}~}$}
\newcommand{\FeAl}{$\mathrm{Fe_{0.98}Al_{0.02}~}$}
\newcommand{\eFeN}{$\epsilon-\mathrm{Fe_{3-x}N~}$}
\newcommand{\eFeyN}{$\epsilon-\mathrm{Fe_{2.28}N~}$}

\begin{document}

\title{Formation of iron nitride thin films with Al and Ti additives}

\author{Rachana Gupta}

\affiliation{Institute of Engineering and Technology, Devi Ahilya
Vishwavidyalaya, Khandwa Road, Indore - 452 017, India}

\author{Akhil Tayal} \author {Mukul Gupta} \email{mgupta@csr.res.in} \author {Ajay Gupta}

\affiliation{UGC-DAE Consortium for Scientific Research,
University Campus, Khandwa Road, Indore - 452 001,India}

\author{M. Horisberger}

\affiliation{Laboratory for Developments and Methods, Paul
Scherrer Institut, CH-5232 Villigen PSI, Switzerland}

\author{J. Stahn}

\affiliation{Laboratory for Neutron Scattering, Paul Scherrer
Institut, CH - 5232 Villigen PSI, Switzerland}

\date{\today}


\begin{abstract}

In this work we investigate the process of iron nitride (Fe-N)
phase formation using 2\pat Al or 2\pat Ti as additives. The
samples were prepared with a magnetron sputtering technique using
different amount of nitrogen during the deposition process. The
nitrogen partial pressure (\pn) was varied between 0-50\% (rest
Argon) and the targets of pure Fe, [Fe+Ti] and [Fe+Al] were
sputtered. The addition of small amount of Ti or Al results in
improved soft-magnetic properties when sputtered using \pn $\leq$
10\p. When \pn is increased to 50\p~ non-magnetic Fe-N phases are
formed. We found that iron mononitride (FeN) phases (N\pat
$\sim$50) are formed with Al or Ti addition at \pn =50\p~whereas
in absence of such addition \eFeN phases (N\pat$\sim$30) are
formed. It was found that the overall nitrogen content can be
increased significantly with Al or Ti additions. On the basis of
obtained result we propose a mechanism describing formation of
Fe-N phases Al and Ti additives.

\end{abstract}



\maketitle


\section{Introduction}
\label{1}

Reactive sputtering of iron with nitrogen results in
nanocrystallization and amorphization of the deposited films.
\cite{Hellgren_PR99,Babonneau_APL03,Gupta_PRB02,Gupta_JJPS04,Gupta:PRB05}
This happens due to incorporation of nitrogen atoms at the
interstitial sites resulting in expansion and distortion of bcc-Fe
units cells. This essentially implies that reactive nitrogen
sputtering plays an important role in refining the grain sizes. As
the grain size decreases below the ferromagnetic exchange length
($\sim$20\,nm for Fe) the averaging is done on very fine grains
and magnetization follow the easy direction of each individual
grain. This results in very low coercivity. Further, as the grain
size decreases, there is a significant increase in the volume
fraction of grain boundaries or interfaces. Therefore, the value
of saturation magnetization is reasonably high. A combined effect
results in good soft-magnetic properties in nanocrystalline or
amorphous iron nitrides. However, incorporation of nitrogen in
iron also results in loss of magnetization due to formation of non
magnetic Fe-N compounds due to covalent Fe-N bonds. This leads to
magnetic anisotropy and an increase in coercivity. In addition it
has been observed in magnetic Fe-N that and nitrogen atoms tend to
diffuse out even at very low temperatures.
~\cite{Navio.NJP2010,Menendez.JAP01,Gupta:AM:2009} Therefore
nanocrystalline iron nitrides have these intrinsic limitations in
succeeding as good soft-magnetic alloys.

Since early 1990s to recent years, various Fe-N thin films were
studied by adding a tiny amount of a third element say $X$, it was
observed that using such additives the limitations of binary Fe-N
can be reduced to a great
extent.~\cite{Hasegawa:JMSJ:1990,Ishiwata:JAP:1991,Ono:JAP:1993,Takeshima:JAP:1993,
Qiu:JAP:1994,Roozeboom:JAP:1995,Viala:JAP:1996,Wang:JPCM:1997,Varga:JAP:1998,
Chen:JAP:2000,Liu:APL:2000,Chezan:PSSA:2002,Rantschler:JAP:2003,Liu:JAP:2003,Chechenin:JPCM:2003,Kazmin:TPL:2005,
Das:PRB:2007,Sangita:PRB:2008,FengXu:JAP:2011} The basic idea
behind adding such element $X$ was to chose an element which has
more affinity to nitrogen than iron. When a third element $X$ =
Al, Ti, Ta, Zr etc. is added in an appropriate amount it can be
substitutionally dissolved in to bcc-Fe unit
cell.~\cite{Das:PRB:2007} This addition results in increased
thermal stability of the system as the binding energy of the
system increases because the heat of formation of $X$-N is less
than Fe-N.~\cite{Viala:JAP:1996} Therefore element $X$ works like
a trap for nitrogen diffusing out. Although many third elements
have been used in Fe-$X$-N system, the choice of element $X$ has
been rather arbitrary. For example earlier works were mainly
focused on Fe-Ta-N
systems~\cite{Ishiwata:JAP:1991,Takeshima:JAP:1993,Qiu:JAP:1994,Viala:JAP:1996,Varga:JAP:1998},
more recently other elements e.g.
Ti~\cite{Wang:JPCM:1997,Ding:JAP:2002,Rantschler:JAP:2003,Das:PRB:2007},
Al~\cite{Liu:JAP:2003,Kazmin:TPL:2005},
Zr~\cite{Chezan:PSSA:2002,Chechenin:JPCM:2003} and
Rh~\cite{Chen:JAP:2000} etc. have been used.

In this work we have chosen $X$ = Al, Ti in Fe-$X$-N, as the
atomic radii $ar\mathrm{_{Fe}}$ = 0.156\,nm is larger than
$ar\mathrm{_{Al}}$ = 0.118\,nm but smaller than $ar\mathrm{_{Ti}}$
= 0.176\,nm.~\cite{Clementi:JCP:1967} Therefore, Ti addition is
expected to expand the units cell of Fe when substitutionally
dissolved whereas no such expansion is expected with Al addition.
By adding similar amount of Ti and Al i.e. 2\pat we systematically
studied formation of Fe-N phases. The addition of small amount of
Al or Ti has been reported to increase the thermal stability of
Fe-N.~\cite{Liu:APL:2000,MSEB:2003:Ma,Wang:JPCM:1997,JPCM:Jiang:1997}
It may be noted that though the heat of formation of TiN and AlN
is almost similar,~\cite{Kopcewicz:JAP:1995} their affinity to N
is different.~\cite{Viala:JAP:1996} Therefore the mechanism by
which addition of element $X$ affects formation of Fe-N phases is
not very clear. In addition, the studies with addition of element
$X$ in Fe-$X$-N system have been mainly focused to magnetic phases
where small amount of reactive nitrogen is used during sputtering.
In the present work we investigate for the first time the effect
of Al or Ti addition on formation of non-magnetic iron nitride
phases. Recently, non-magnetic iron mononitrides have emerged as a
promising material in spintronics
applications.~\cite{Wit:PRL94,MG:JAC2001,Navio.PRB08,Navio:APL:2009}
A controlled annealing of FeN produces the
$\gamma^{\prime}$-Fe$_4$N phase and thus provides a source of spin
injection for semiconductors or diluted magnetic
semiconductors~\cite{Gallego.PRB04}.

In our approach we have taken extreme care to deposit a series of
samples in a single sputtering run so as to ensure identical
depositions conditions. The amount of reactive nitrogen was
successively increased from 0-50\% with a step of 5\%. The
structure of deposited phases was studied using x-ray diffraction.
In order to measure magnetic moment of the samples precisely, we
used polarized neutron reflectivity. The magnetization
measurements were carried out using a SQUID magnetometer and
conversion electron M\"{o}ssbauer spectroscopy.

\section{Experimental}
\label{2}

Thin film samples were prepared at room temperature using a direct
current magnetron sputtering (dc-MS) deposition system. The
samples were prepared simultaneously on Si\,(100) and float glass
substrates. A mixture of argon and nitrogen gases at different
ratio was used to sputter a target. The total gas flow was kept
constant at 20\,standard cubic centimeter per minute (sccm). The
nitrogen partial pressure defined as \pn =
$P\mathrm{_{N_2}}$/($P\mathrm{_{Ar}}$+$P\mathrm{_{N_2}}$) was
varied at 0, 5, 10, 15, 20, 25, 30, 35, 40 and 50\p. A base
pressure of $\approx1\times10^{-7}$\,mbar was achieved prior to
deposition. During the deposition the partial pressure in the
chamber was $\approx4\times10^{-3}$\,mbar. The targets of pure Fe,
and composite targets of [Fe+Ti], and [Fe+Al] were sputtered using
a power of 50\,W. Before deposition the vacuum chamber was
repeatedly flushed with argon and nitrogen gases so as to minimize
the possible contamination of trapped gases inside the vacuum
chamber. The targets were pre-sputtered at least for 10\,minutes
in order to remove possible surface contaminations.

\begin{figure} \center
\includegraphics [width=90mm,height=75mm] {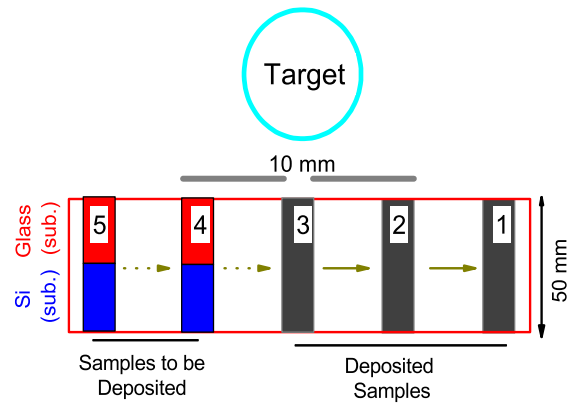}
\caption{\label{fig:schema} (Colour online) Schematic
representation of target and substrate configuration used for
depositing thin film samples in dc-MS deposition system.}
\end{figure}

In order to achieve similar deposition conditions for a set of
samples i.e. sputtering of Fe, [Fe+Ti] or [Fe+Al] targets all
parameters were kept constant, except \pn. On a 99.95\p~pure
target of Fe, small pieces of 99.999\p~pure Al or 99.99\p~Ti were
pasted using a silver epoxy paste. Thin films of different \pn
were prepared without exposing the deposition chamber to the
atmosphere. The targets were covered with a small slit of size
about 10\,mm and the substrate was exposed at center of the
target. After deposition at a particular \pn, the substrate was
translated using a computer controlled linear translation stage.
Once deposition of a sample at a particular \pn is completed, the
gas flows were changed using mass flow controllers and monitored
using a residual gas analyzer until a stable gas ratio is
obtained. Such a procedure minimized the variation in deposition
conditions which might occur otherwise. A schematic diagram of
target and substrate configuration is shown in
fig.~\ref{fig:schema}. A total of 10 samples were deposited at
various \pn in a series. On a glass substrate of 250\,mm,
Si\,(100) substrates were mounted on position 1 to 10
(schematically shown in fig.~\ref{fig:schema} for positions 1 to
5). The thickness of the films was calibrated with x-ray
reflectivity and by using different deposition times it was kept
at about 100\,nm for all samples prepared in this work.

\begin{figure*} \center
\includegraphics [width=85mm,height=80mm] {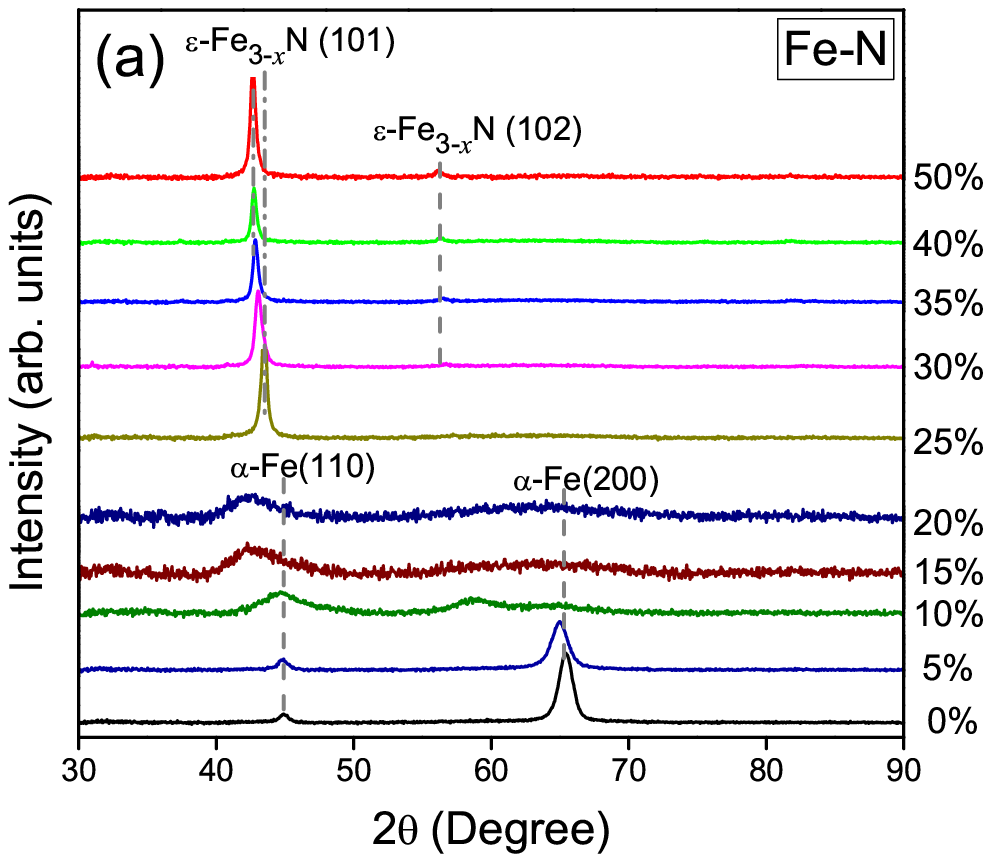}
\includegraphics [width=85mm,height=80mm] {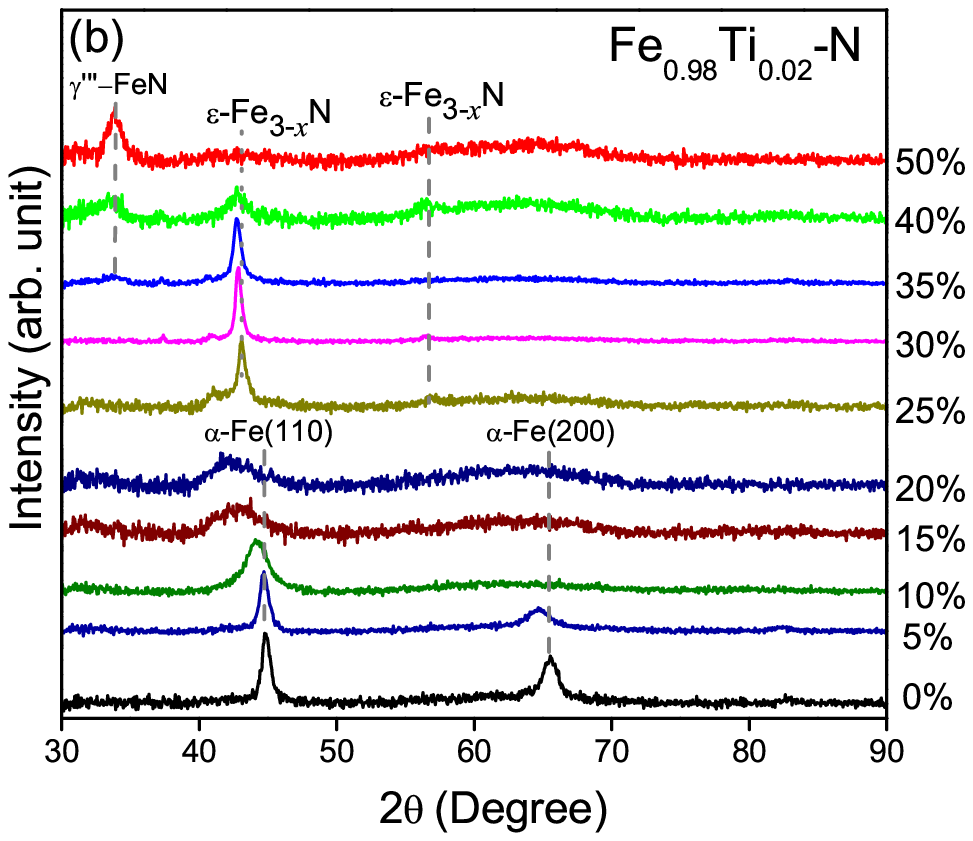}
\includegraphics [width=85mm,height=80mm] {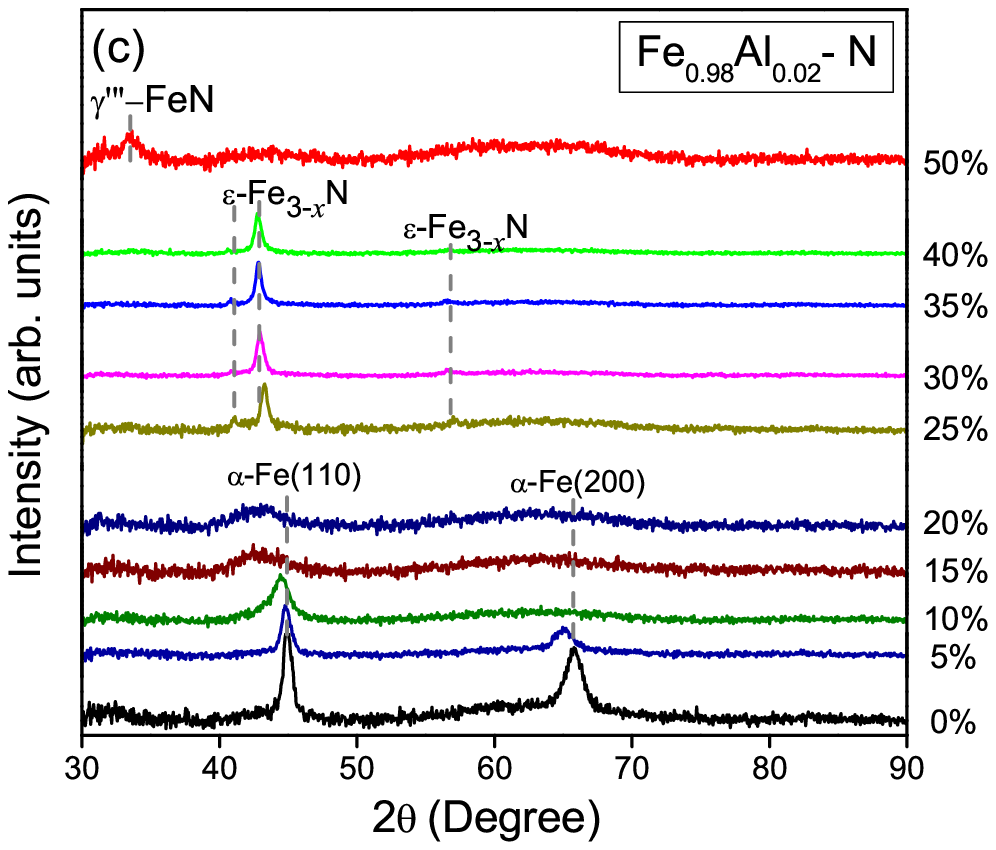}
\includegraphics [width=80mm,height=70mm] {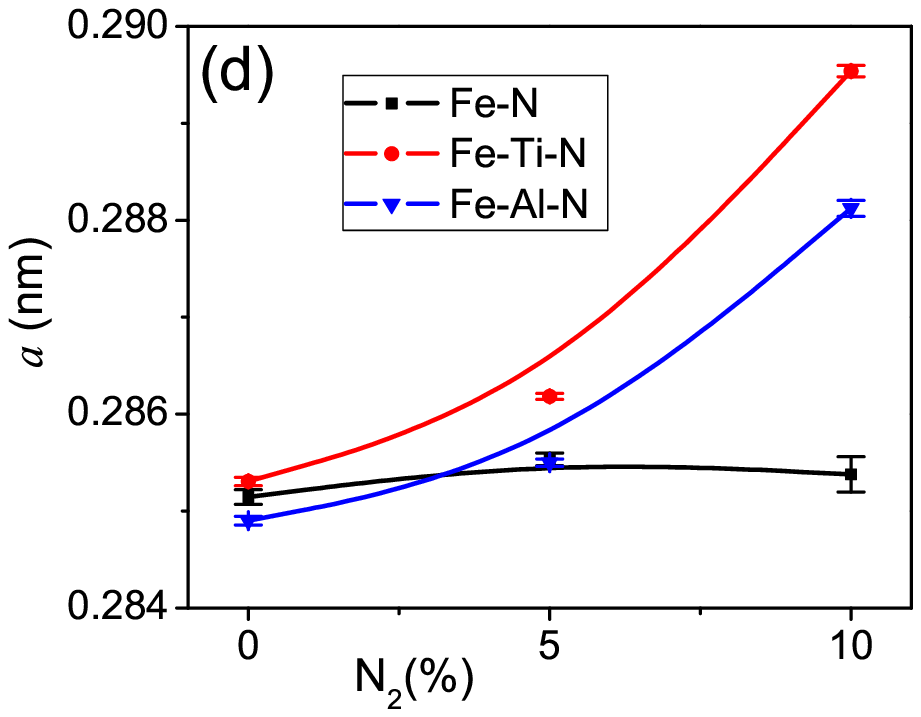}
\caption{\label{fig:FeN_XRD} (Colour online) X-ray diffraction
pattern of iron nitride (Fe-N) thin films (a), \FeTi-N films (b)
and \FeAl-N thin films (c) prepared using nitrogen partial
pressure, \pn = 0--50\% (rest Argon). The lattice parameter
obtained from XRD data for samples prepared for \pn = 0, 5 and
10\% are plotted in (d), here solid lines are guide to eye.}
\end{figure*}

The composition of deposited Fe-Ti and Fe-Al films (prepared
without nitrogen) was determined using energy dispersive x-ray
analysis installed in a scanning electron microscope. The obtained
composition are \FeTi and \FeAl. The structural characterizations
of the samples were carried using x-ray diffraction (XRD) using a
standard x-ray diffractometer (Bruker D8 Advance) equipped with
Cu\,K-$\alpha$ x-rays in $\theta$-2$\theta$ mode. A 1-D position
sensitive detector based on silicon strips technology (Bruker
LynxEye) was used. By using using such detector it was possible to
obtained XRD data with good statistics. The magnetization
measurements were carried out using a Quantum design SQUID
magnetometer and polarized neutron reflectivity (PNR). The PNR
measurements were performed at (AMOR)~\cite{Gupta_PramJP04} and
NARZISS neutron reflectometers at SINQ/PSI, Switzerland. The
Conversion electron M\"{o}ssbauer spectroscopy (CEMS) measurements
were carried out at room temperature using a \che{^{57}Co} source
embedded in a rhodium matrix. The conversion electrons were
detected by a proportional counter having continuous flow of a
helium-methane (5\p~ methane balance helium) gas mixture.

\section{Results}
\label{3}

\subsection{X-ray diffraction} \label{3.1}

Fig.~\ref{fig:FeN_XRD} shows the XRD pattern of Fe-N samples
prepared using \pn = 0-50\p. The obtained structural changes are
similar as observed in an earlier work.~\cite{Gupta:PRB05} Here,
samples without Al or Ti additions were essentially prepared as a
reference so that effect of alloying elements can be compared
precisely. However, as we use a small slit of 10\,mm to control
over the deposition area there are some interesting features we
observed in the XRD patterns shown in fig.~\ref{fig:FeN_XRD} (a)
-(c). It may be noted that nitrogen free (\pn = 0\p) sample have a
bcc-$\alpha$-Fe structure with preferred orientation along (200)
direction. Normally pure iron thin films are expected to align in
(110) direction. The preferred orientation along (200) direction
was caused by the small slit placed beneath the target as shown in
fig.~\ref{fig:schema}. Such slit was necessary to control
deposition area so as to deposit ten samples with different \pn.
This slit basically reduced the number of adatoms depositing on a
substrate. In this situation preferred orientation may change from
(110) to (200) direction as the surface energy of (200) plane is
lower than that of (110) plane. However, when no such slit is used
the large influx of adatoms may give rise to some strain which
results in preferred orientation along (110) direction. Such
behaviour with sputtering power has been observed in case of
Ti~\cite{Oh:JAP:93} and with substrate biasing in case of
Cr~\cite{Feng:JAP:94}. However, as we observe from
fig.~\ref{fig:FeN_XRD} (a) with reactive sputtering, the preferred
orientation seems to change from (200) to (110) direction. This
effect is more prominent when samples were prepared with addition
of Al or Ti. As discussed later in section~\ref{4}, reactive
sputtering with nitrogen or with Al or Ti additions, the deposited
films cause strain in the Fe and in this case the strain energy
may dominate over the surface free energy.

\begin{table}
\caption{\label{tab:tab:grainsize} Average grain size of samples
prepared using different nitrogen partial pressure.}
\begin{ruledtabular}
\begin{tabular}{cccc}
\pn&\FeN&\FeTi-N&\FeAl-N\\ \hline
0\p&9.0(0.3)&11.3(0.3)&10.3(0.3)\\ \hline
5\p&7.8(0.2)&11(0.3)&9.6(0.3)\\ \hline
10\p&2.4(0.3)&3.8(0.1)&4.1(0.1)\\ \hline
15\p&amorphous&amorphous&amorphous\\
\end{tabular}
\end{ruledtabular}
\end{table}

We will first analyze the XRD pattern of samples prepared without
Al or Ti additions as shown in fig.~\ref{fig:FeN_XRD} (a). Here
the line width of the diffracted pattern can be used to calculate
the grain size of the diffracting specimen in the direction
perpendicular to the plane of the film using Scherrer
formula,\cite{Cullity_XRD,Scherrer_Eq:NatureNano:2011} $t$ =
0.9$\lambda$/$b$ $\cos$ $\theta$, where $t$ is the grain size, $b$
is an angular width in terms of 2$\theta$, $\theta$ is the Bragg
angle and $\lambda$ is the wavelength of the radiation used.
Although the boundary between nanocrystalline and amorphous phases
may not be sharp, the obtained grain sizes (see
table~\ref{tab:tab:grainsize}) may be used to identify the
structure as nanocrystalline or amorphous. For pure iron sample
the grain size calculated using (110) reflection is 9.0\,nm which
decreases slightly to 7.8\,nm when sputtered using \pn = 5\p. When
\pn is increased to 10\p, the estimated grain size is about
2.4\,nm therefore at this partial pressure it may be difficult to
label the structure as nanocrystalline or amorphous. Above this
\pn, the peaks are broad enough to identify them as amorphous.
This amorphous phase persists up to nitrogen partial pressure of
20\p~ and for \pn $\geq$ 25\p, \eFeN compounds are obtained. Since
the heat of formation for \eFeN phases is about (-40 to
-45\,\che{kJ~mol^{-1}} as compared with neighboring, e.g.,
\che{Fe_{4}N} (-12\,\che{kJ~mol^{-1}}) or \che{Fe_{2}N}
(-34\,\che{kJ~mol^{-1}}), phases.~\cite{Tessier_SSS00} Therefore
from the energetics of binary iron nitrides at room temperature it
is expected that \eFeN phase should be readily formed as enough
reactive nitrogen is made available which may be the case when \pn
exceeds 25\p.

On the basis of amount of nitrogen partial pressure used, the
phases formed can be divided into three ranges: (i) \pn $\leq$
10\p,~ where nanocrystalline Fe-N phases are formed (ii) \pn = 10
- 20\p, where amorphous Fe-N phase are obtained and (iii) \pn
$\geq$ 25\p, where \eFeN or \gt-\FeN phases (with addition of Al
or Ti) are obtained. We will now discuss the effect of small
amount of Al or Ti in formation of iron nitrides in these three
\pn ranges.

\begin{itemize}

\item[(i)] Fig.~\ref{fig:FeN_XRD}(b) and fig.~\ref{fig:FeN_XRD}(c)
shows the XRD pattern of samples with addition of Al and Ti,
respectively for different \pn. For nitrogen free samples, the
average grain size is about 9\,nm (see
table~\ref{tab:tab:grainsize}) which increases slightly to
11.3\,nm with Ti addition and 10.3\,nm with Al addition. At \pn =
5\p,~the grain sizes decreases slightly both with Al or Ti
additions. At \pn = 10\p,~the grain sizes decrease appreciably to
about 4\,nm. The positions of Bragg peaks shift to lower-angle
side when \pn is increased from 0-10\p~ although the basic
structure remains same. This happens as nitrogen atoms occupy
interstitial sites within the bcc unit cell. The lattice parameter
($a$) can be calculated for (110) peak and is plotted in
fig.~\ref{fig:FeN_XRD} (d). As can be seen here with Al or Ti
additions $a$ increases more significantly. The reasons for such
behaviour will be discussed later in section \ref{4}.

\item[(ii)] For \pn = 15 and 20\p,~amoprhous phase are obtained
and here no appreciable effect of Al or Ti addition can be seen in
the XRD pattern.

\item[(iii)] For \pn between 25-35\p,~the results are similar to
Fe-N case where \eFeN compounds are obtained. However at \pn = 40
and 50\p~, the results are markedly different. With Ti addition as
can be seen in fig.~\ref{fig:FeN_XRD} (b) a new phase start
appearing for \pn = 40\p,~where as with Al addition this phase is
visible only at \pn = 50\p. This new phase can be indexed as
\gt-FeN (confirmed with CEMS measurements shown later) which is
iron mononitride and generally formed when iron targets are
sputtered using
100\p~nitrogen.~\cite{gupta:PRB02,Gupta:PRB05,MG:JAC:2011}
Formation of this high nitrogen phase with Al or Ti addition can
be understood with an enhancement of nitrogen incorporation.
Although \gt-FeN phase can be obtained when sputtered only with
nitrogen as sputtering gas, formation of this phase with a
50\p~Ar+50\p~N$_2$ mixture with Al or Ti addition is interesting
as deposition rates will much higher as compred to the case when
sputtering is done using N$_2$ alone as sputtering gas. The
\gt-FeN phase has lowest heat of formation
(-47\,\che{kJ~mol^{-1}}) therefore due to enhancement of nitrogen
by adding Ti or Al in Fe results in formation of \gt-FeN phase.
These results will be discussed in detail in section \ref{4}.

\end{itemize}

\begin{figure} \center
\includegraphics [width=42mm,height=55mm] {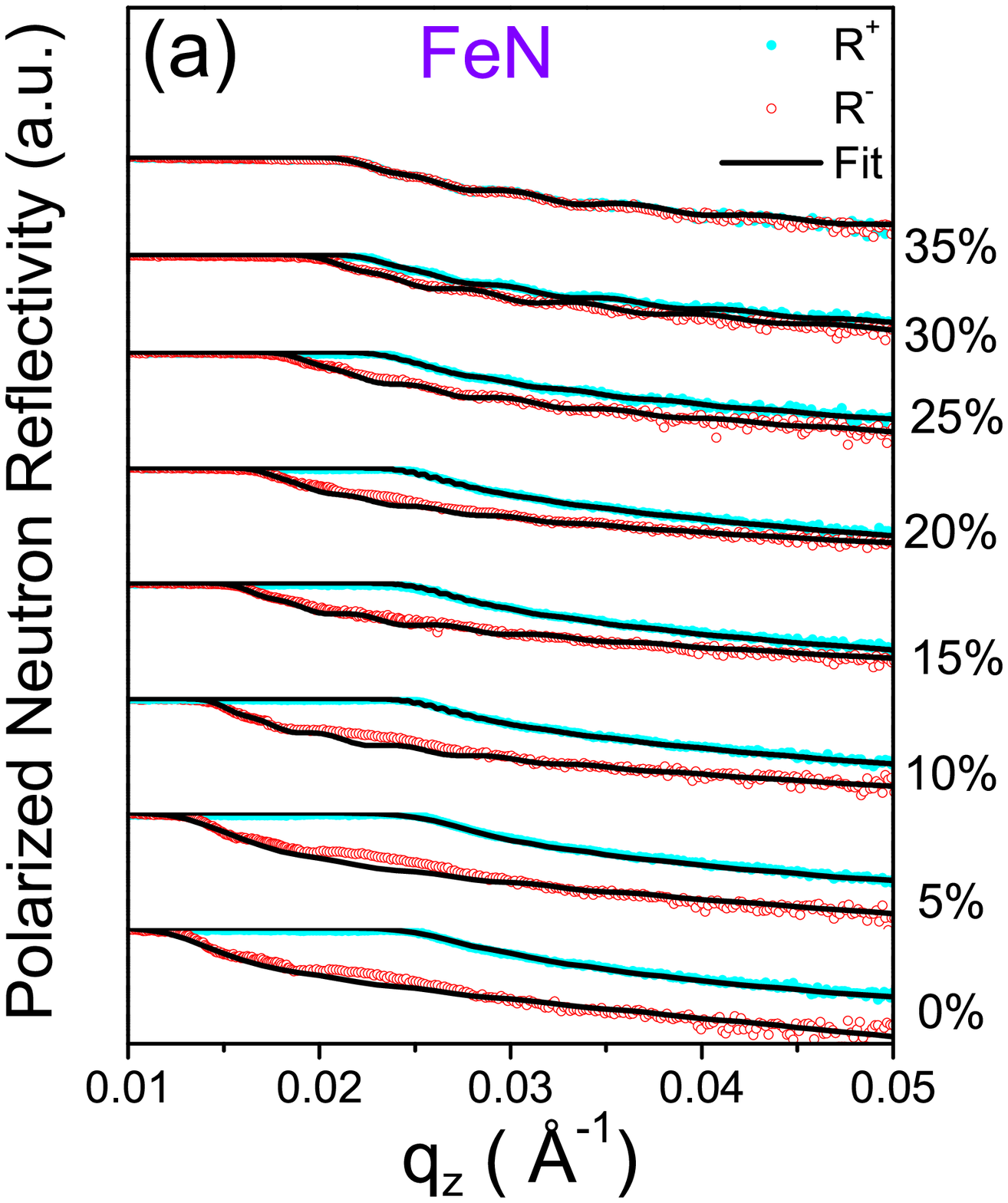}
\includegraphics [width=42mm,height=55mm] {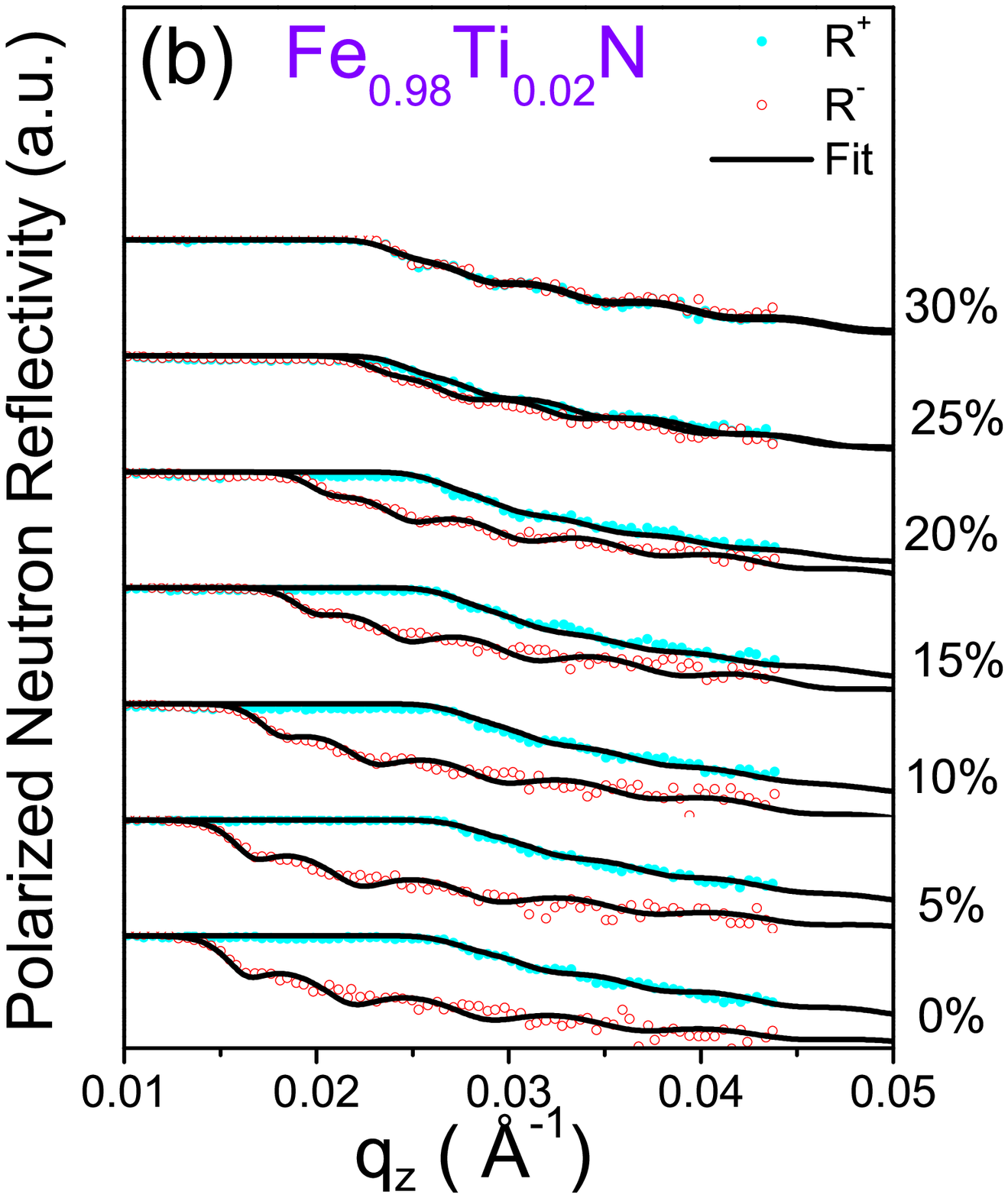}
\includegraphics [width=42mm,height=55mm] {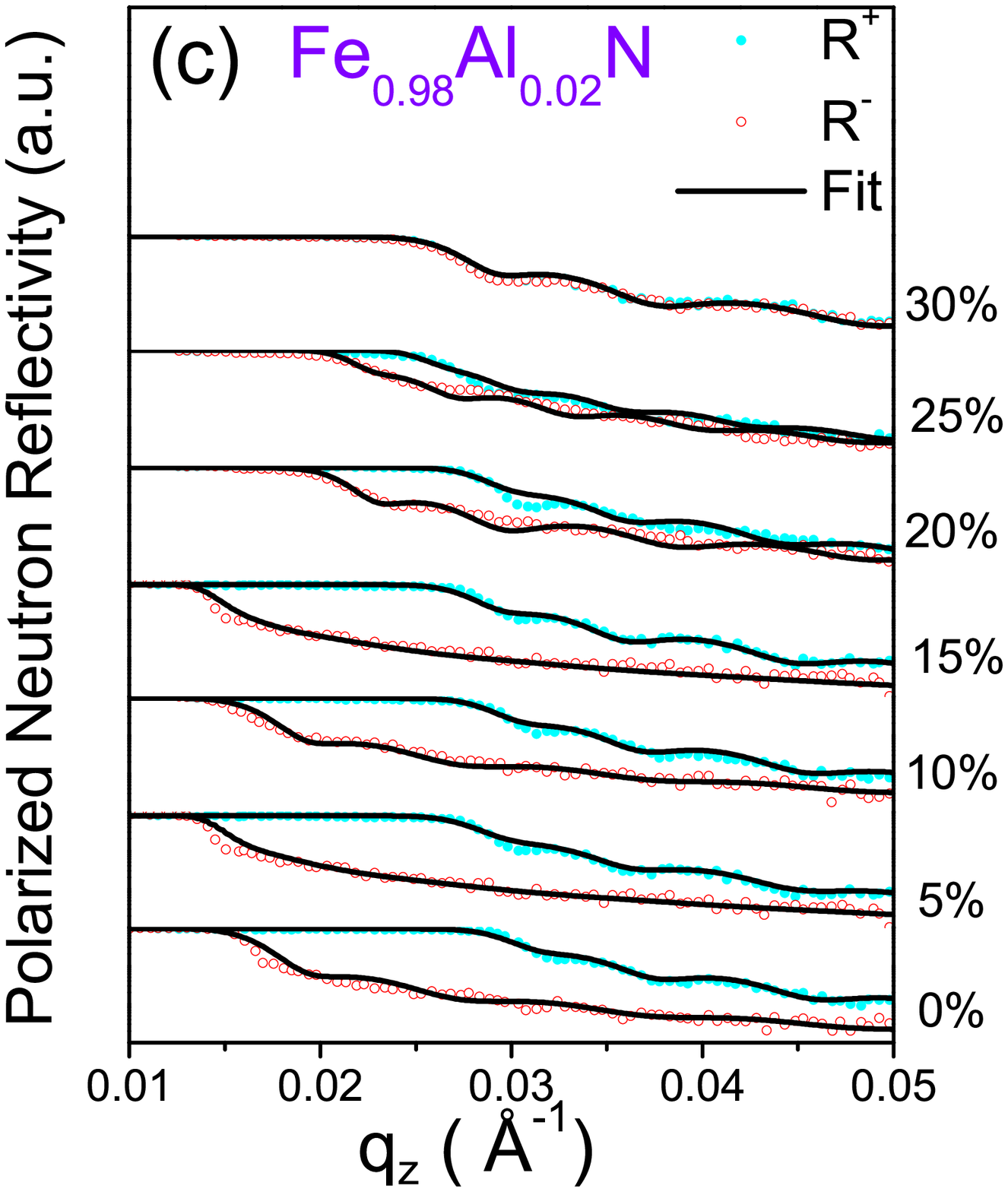}
\includegraphics [width=43mm,height=47mm] {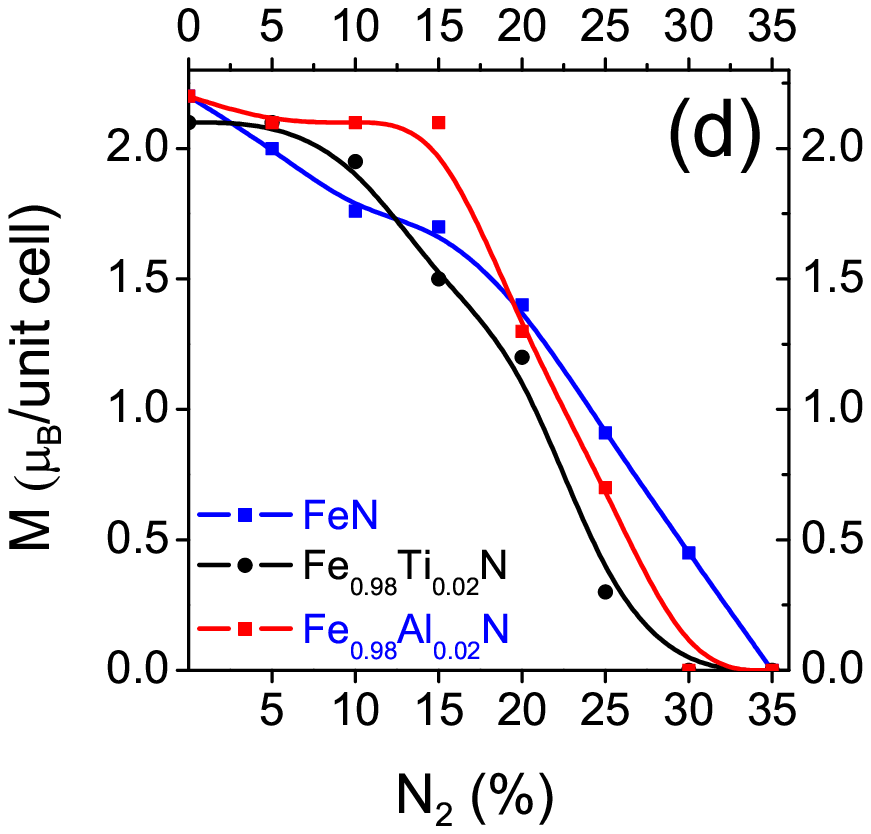}
\caption{\label{ALL_PNR} (Color online) Polarized neutron
reflectivity of Fe-N (a) \FeTi-N (b) and \FeAl-N (c) samples as a
function of nitrogen partial pressure. The scattered points
correspond to experimental data and solid lines are fit to them.
The reflectivity pattern are vertically shifted for clarity. A
decay of magnetic moment for \FeN, \FeAl-N and \FeTi-N samples
with different \pn is shown in (d). Here scattered points are
measured data and solids line are guide to eye.}
\end{figure}

\begin{figure} \center
\includegraphics [width=85mm,height=85mm] {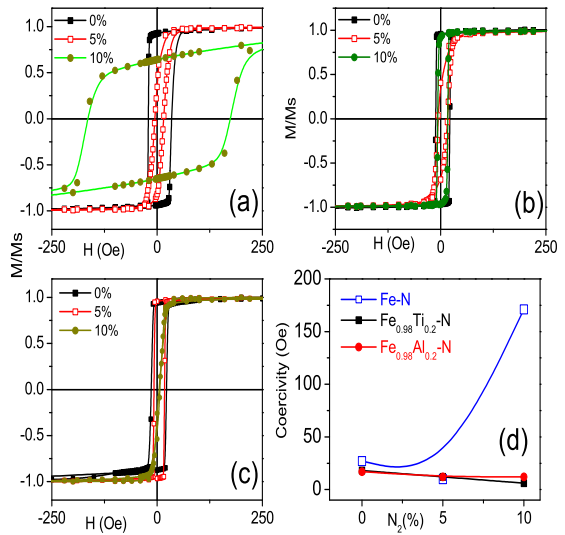}
\caption{\label{MH} (Color online) Magnetization measurements of
Fe-N (a), \FeTi-N(b) and \FeAl-N(c) thin films prepared using \pn
= 0, 5, and 10\p. A variation of coercivity as a function of \pn
(d). Here scattered points correspond to experimental data and
solid line are guide to eye.}
\end{figure}

\subsection{Polarized neutron reflectivity measurements} \label{3.2}

Polarized neutron reflectivity (PNR) is a technique which is able
to yield the absolute value of magnetic moment per atom in a
magnetic thin film with high accuracy.\cite{Blundell_PRB92} In
contrast to bulk magnetization magnetometer technique (e.g. DC
extraction, VSM or SQUID), no correction due to diamagnetic signal
from the substrate has to be applied in PNR. Further, the samples
dimensions and mass does not play any role in determination of
magnetic moment. During the experiment, polarized neutrons with
spin parallel or antiparallel to the direction of magnetization on
the sample are reflected-off the surface of the sample at grazing
incidence. The measurements were performed with an applied field
of 800\,Oe, which is sufficient to reach the saturation
magnetization in all the samples.

Figure~\ref{ALL_PNR} shows PNR pattern of Fe-N, \FeTi-N and
\FeAl-N samples prepared using different \pn. As the amount of
nitrogen partial pressure is increased, the critical edge in spin
down reflectivity ($R^{-}$) shows a shift towards higher q$_{z}$
values, and the separation between $R^{+}$ and $R^{-}$
reflectivities decreases continuously. For Fe-N samples $R^{+}$
and $R^{-}$ converge for \pn = 35\p~, whereas in case of \FeAl-N
and \FeTi-N samples the two reflectivities converge at 30\p.

The difference between spin up and down reflectivities is a
measure of magnetic moment. The PNR profiles were fitted using a
computer program\cite{SimulReflec} based on Parratt's
formulism.~\cite{Parratt.PR54} The obtained values of iron
magnetic moment ($\mathrm{\mu_{B}}$) per atom are plotted in
fig.~\ref{ALL_PNR}(d) as a function of \pn. As can be seen here,
the magnetic moment decays as nitrogen is added is the system. Up
to \pn = 10\p, this decay is faster when Al or Ti are not added.
However, for \pn $\geq$ 20\p, the decay in magnetic moment is
faster with addition of Al or Ti. This result can be understood in
terms of formation of Fe-N bonds which are readily formed when
additives Al or Ti are not added. As discussed later in section
~\ref{4}, with Al or Ti addition nitrogen essentially interacts
with Al or Ti and therefore the magnetic moment of Fe does not
decay as fast as it will be when Al or Ti are not added. When \pn
is further increased the total amount of nitrogen that can be
added into the system is much exceeds due to interaction of Al or
Ti with nitrogen. This leads to a faster loss of magnetization in
samples prepared with Al or Ti addition for \pn $\geq$ 15\p. When
\pn exceeds 30\p, the spin up and down reflectivities remain
identical indicating that samples are not ferromagnetic.

\subsection{Magnetization measurements} \label{3.3}

The magnetization measurements on different samples were carried
out using a SQUID magnetometer. Fig.~\ref{MH} shows normalized M-H
loops for Fe-N, \FeTi-N and \FeAl-N samples prepared using \pn =
0, 5 and 10. The obtained values of coercivity ($\mathrm{H_c}$)
are plotted in fig.~\ref{MH} (d). For nitrogen free samples, the
value of $\mathrm{H_c}$ for Fe thin film is about 30\,Oe, which
decreases to about 17\,Oe, when Al or Ti added. When sputtered
using 5\p~ nitrogen, $\mathrm{H_c}$ decreases to about 10\,Oe. It
may be noted that when sputtered using 5\p~ nitrogen the average
grain decreases (see table~\ref{tab:tab:grainsize}). A reduction
in grain size below ferromagnetic exchange length allows exchange
coupling between the neighboring grains and results in a reduced
effective
anisotropy.~\cite{Gupta:PRB05,Herzer_IEEE90,Herzer_IEEE89} The
results obtained for the samples which were prepared using \pn =
10\p~ are very interesting. For the sample prepared without Al or
Ti additions, $\mathrm{H_c}$ increases significantly to 170\,Oe
whereas the samples prepared using Al or Ti it further decreases
to about 5\,Oe. As can be seen from fig.~\ref{ALL_PNR}(d), the
magnetic moment decreases rather slowly with Al or Ti additions.
Therefore with Al or Ti additions, soft-magnetic properties can be
improved. The obtained results can be understood in terms of
structural changes caused by reactive sputtering. When Fe was
sputtered with 10\p~ nitrogen, formation of non-magnetic Fe-N
covalent bonds may introduce large anisotropy in the system as
reported in literature.~\cite{Gupta:PRB05} Whereas in presence of
Al or Ti alloying elements, nitrogen predominantly interacts with
these alloying elements and due to formation of finer grain
nanocrystalline structure, the coercivity decreases further.

\begin{figure} \center
\includegraphics [width=85mm,height=120mm] {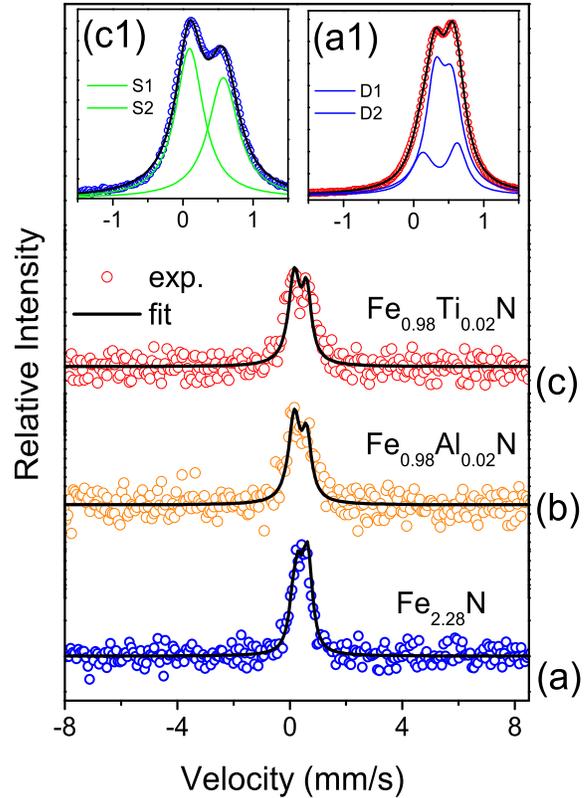}
\vspace {-5mm} \caption{\label{cems} (Colour online) Conversion
electron M\"{o}ssbauer spectroscopy measurements on Fe-N (a),
\FeAl-N (b) and \FeTi-N (c) and  samples prepared using \pn =
50\p. Inset shows CEMS pattern of samples prepared with $^{57}$Fe
and recorded at a reduced velocity for Fe-N (a1) and for \FeTi-N
(c1).}
\end{figure}

\begin{table} [!hb]
\caption{\label{tab:tab:CEMS} Conversion electron M\"{o}ssbauer
spectroscopy parameters: isomer shift (IS), quadrupole shift (QS)
and relative area (RA) for Fe-N, \FeTi-N and \FeAl-N samples
prepared with \pn = 50\% nitrogen.}
\begin{ruledtabular}
\begin{tabular}{cccc}
Parameter&\FeN&\FeTi-N&\FeAl-N\\ \hline
IS1 ($\pm$ 0.02)&0.45&0.01&0.1\\
IS2 ($\pm$ 0.02)&0.38&0.64&0.66\\ \hline
QS1 ($\pm$ 0.04)&0.25&--&--\\
QS2 ($\pm$ 0.04)&0.50&--&--\\ \hline
RA1 ($\pm$ 3\%)&63&54&62\\
RA2 ($\pm$ 3\%)&37&38&46\\
\end{tabular}
\end{ruledtabular}
\end{table}

\subsection{Conversion electron M\"{o}ssbauer spectroscopy measurements} \label{3.4}

From XRD measurements we observed that at the highest nitrogen
partial pressure used in this work, new phases of Fe-N are
obtained with Al or Ti addition. Therefore in order to understand
the phase formed with Al or Ti addition, we performed conversion
electron M\"{o}ssbauer spectroscopy (CEMS) measurements in the
samples prepared using \pn = 50\p. The measurements were first
performed with velocity of the drive between $\pm$8.5\,mms$^{-1}$,
which is sufficient to cover all resonance lines of magnetic Fe.
Fig.~\ref{cems} shows the CEMS pattern corresponding to Fe-N,
\FeAl-N and \FeTi-N samples prepared with 50\pat of \pn. The CEMS
pattern of the Fe-N samples was found to have asymmetric doublets
and no magnetic lines can be seen. In order to get precise
information about the samples we prepared Fe-N and \FeTi-N samples
using $^{57}$Fe enrichment and the CEMS measurements were
performed with a reduced drive velocity between $\pm$2\,mms$^{-1}$
to get better CEMS data. The inset of fig.~\ref{cems} shows CEMS
pattern of $^{57}$Fe-N and $^{57}$\FeTi-N samples. The patterns
with Ti addition are clearly different. From XRD results we find
that the Fe-N phase formed has \eFeN type structure. Therefore in
order to fit the observed symmetric doublets we deconvoluted it
into two doublets corresponding to Fe-III and Fe-II sites. The
fitted parameters are given in table II. These value matches well
with reported values for \eFeN and using the relative area ratio,
we obtain the value of x = 0.72 following a procedure given in
ref.[~\cite{Schaaf_HypInt95}]. This gives the composition of the
sample as \eFeyN.

On the other hand CEMS pattern of samples prepared with Al or Ti
addition were different as compared to a sample prepared without
such additions. Here it was found that the CEMS pattern is rather
more asymmetric. The XRD pattern also revealed that the crystal
structure is completely different and correspond to ZnS-type fcc
structure. The phase identified was a monoatomic iron nitride
having 1:1 atomic ratio Fe and N. As described in the literature,
fitting of such patterns can be done considering two singlets. The
fitted parameters are given in table II and match well with the
reported values.~\cite{Borsa.HI.2003,MG:JAC:2011} Here the
singlets with almost zero value of isomer shift corresponds to Fe
coordinated tetrahedrally with four N atoms and singlets with
higher value of isomer shift corresponds to defects or vacancies.
Therefore it can be seen that with Al or Ti addition Fe-N phase
having more nitrogen can be obtained.

In an earlier work Liu et al~\cite{Liu:JAP:2003} investigated CEMS
pattern of Fe-Al-N thin films prepared using 2\pat Al in a wide
\pn range. In their XRD pattern they also observed a peak around
2$\theta\thickapprox$34\,degree, but they identified it as a
``signature of Si$_3$N$_4$". The CEMS pattern of this phase not
presented and most probably they overlooked the formation of iron
mononitride phase.

\section{Discussion}
\label{4}

Combining our results obtained from different experimental
techniques, we investigate a mechanism leading to formation of
iron nitride phases with small addition of Al and Ti. We will
first focus on the samples prepared with highest \pn i.e. 50\p.
Here XRD and CEMS results clearly show that \eFeyN having about
30\pat N is formed when additional element $X$ is not added.
Whereas with Al or Ti addition \gt-\FeN~phase having about 50\pat
N is obtained. Clearly with Al or Ti addition more nitrogen can be
incorporated in iron nitride even though the amount of reactive
nitrogen used during sputtering process remains the same. It can
also be seen from XRD pattern that high-N \gt-\FeN~phase already
start appearing at \pn=40\p when Ti is added where with Al
addition only \eFeN is obtained. This indicates that in Ti
addition is more effective than Al in increasing nitrogen content
in \FeN. This behavior can also be observed in the PNR data where
the magnetic moment falls-off more rapidly with Ti addition as
compared to Al addition for \pn$\geq$15\p.

\begin{figure} [!hb] \center
\includegraphics [width=75mm,height=40mm] {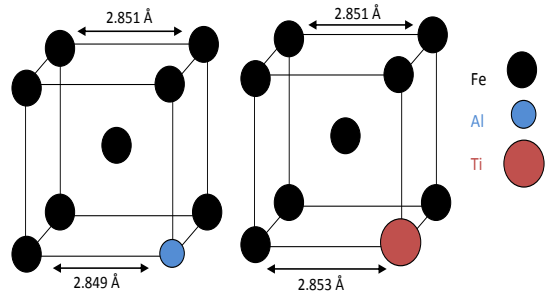}
\caption{\label{CS} (Colour online) Schematic illustration of
bcc-Fe unit cell with Al and Ti additives getting substitutionally
dissolved. The Fe-Fe, Fe-Al, and Fe-Ti distances are obtained from
XRD measurements. }
\end{figure}

As mentioned before, the atomic radii of Fe is larger than Al but
smaller than Ti. Therefore if small amount of Al or Ti is getting
substitutionally dissolved in Fe, it is expected that Ti addition
should expand the unit cell of Fe while Al addition may not be
able todo so. From the XRD measurements performed on samples
prepared without reactive nitrogen, we find that the lattice
parameter ($a$) with Ti addition is 2.853\,{\AA} whereas with Al
addition it is 2.849\,{\AA}. The experimentally obtained value of
$a$ for pure Fe sample is 2.849\,{\AA}. Therefore it appears that
Ti addition is expanding the unit cell slightly while Al addition
is shrinking it slightly. Such a small changes has been reported
in literature with addition of element X.~\cite{Liu:APL:2000} A
schematic representation of this scenario can be understood from
fig.~\ref{CS}. With this argument alone we expect that Ti addition
should result in incorporation of more nitrogen atoms due to
expansion in the unit cell as evidenced in earlier
works.~\cite{Das:PRB:2007} However formation of nearly equiatomic
iron mononitride phase with Al addition can not be understood with
this argument as the unit cell of Fe is actually shrinking with Al
addition. In addition the expansion in unit cell is very small
with Ti addition to cause such an increase in N\pat. Therefore
distortion in the unit cell caused by Al or Ti addition may not
give rise to such a prodigious changes as observed in \pn = 50\%
samples.

\begin{figure} [!hb] \center
\includegraphics [width=75mm,height=65mm] {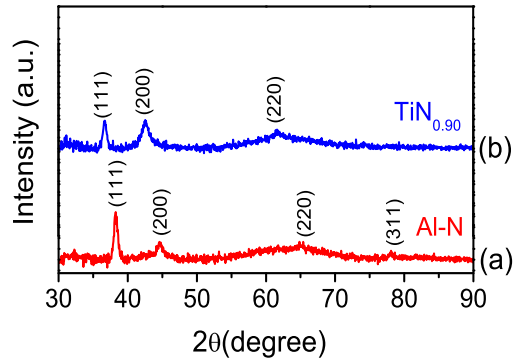}
\caption{\label{XRD:AlN,TiN} X-ray diffraction pattern of Al-N (a)
and Ti-N (b) samples prepared using \pn = 10\p.}
\end{figure}

In order to understand the observed results the affinity of
element $X$ to N and the heat of formation of nitrides should be
looked into. In ternary Fe-$X$-N systems the affinity between
element $X$ and N has been defined with an interaction parameter
$e\mathrm{^X_N}$~\cite{Viala:JAP:1996}. The reported values of
values $e\mathrm{^X_N}$ for Ti and Al are (-0.63) and (+0.0025),
respectively. It may be noted that these values are obtained at a
temperature of about 1600 degree C and their consequences during a
sputtering process may be different.~\cite{Evans:AMIE_Tran:1965}
On the other hand the heat of formation of stoichiometric nitrides
TiN is -337\, kJ mol$^{-1}$ and that of AlN is -320\, kJ
mol$^{-1}$.~\cite{Kopcewicz:JAP:1995} These values are much larger
as compared to FeN which is -47\, kJ
mol$^{-1}$.~\cite{Tessier_SSS00} Therefore it is expected that Ti
or Al should be attracting more nitrogen as compared to iron.

In order to determine the effect of reactive sputtering on Al and
Ti, we prepared a series of Al-N and Ti-N samples under identical
deposition conditions. Here it would be sufficient to compare the
phases formed when Al or Ti are sputtered using \pn = 10\%. The
XRD pattern of Al-N and Ti-N samples prepared using \pn = 10\% are
shown in fig.~\ref{XRD:AlN,TiN}. Here we find that the peaks
corresponding to Ti-N matches well with reported values for
TiN$_{0.91}$ in JCPDS whereas in case of Al-N we find basically
peaks corresponding to fcc-Al. The width of the peaks is larger as
compared to pure Al (not shown). From these results it is apparent
that Ti is indded attracting more nitrogen as compared to Al. In
view of this we can understand formation of nitrogen rich FeN
phase already at \pn = 40\% with Ti addition whereas this phase
shows up at \pn = 50\% with Al addition. From
fig.~\ref{fig:FeN_XRD} (d) it is also clear that as \pn increases
from 0-10\% the lattice parameter of Fe increases more with Ti
addition than with Al addition. In view of above discussion we can
also understand the variation in magnetic moment in the deposited
samples with \pn.

\section{Conclusion}
\label{5}

In the present work we systematically deposited Fe-N thin films
using of small amount (2\p) of Al or Ti as additive elements at
different nitrogen partial pressures. When added in such a small
amount Al or Ti get substitutionally dissolved in Fe. Since Al or
Ti have more affinity to nitrogen as compared to Fe, nitrides of
Al or Ti are formed leading to an enhancement in nitrogen
concentration in Fe-N. It was found that the affinity to nitrogen
is deterministic in overall nitrogen incorporation rather than the
size of additive elements. This was reflected in formation of
non-magnetic iron mononitride phases already at 50\p~ nitrogen
partial pressure which is not possible in absence of additive
elements.

\section*{Acknowledgments} We acknowledge DST for providing financial support to carry out
NR experiments under its scheme `Utilization of International
Synchrotron Radiation and Neutron Scattering facilities'. A part
of this work was performed at the Swiss Spallation Neutron Source,
Paul Scherrer Institute, Villigen, Switzerland.  We are thankful
to Dr.\,P.\,Chaddah for continuous support and encouragement.


%

\end{document}